\documentclass[preprint,12pt]{elsarticle}

\usepackage{amssymb}
\usepackage{graphicx,color}

\journal{J. of Crystal Growth}

\begin{document}

\begin{frontmatter}

\title {Crystal growth of LiGd$_{1-x}$Lu$_x$F$_4$ solid solutions by zone melting technique}


\author[IPEN]{I. A. dos Santos\corref{cor1}}
\cortext[cor1]{Corresponding author}
\ead{iasantosif@usp.br}
\author[IKZ]{D. Klimm}
\author[IPEN]{S. L. Baldochi}
\author[IPEN]{I. M. Ranieri}

\address[IPEN]{Instituto de Pesquisas Energ\'eticas e Nucleares, CP 11049, Butant\~a 05422-970, S\~ao Paulo, SP, Brazil}
\address[IKZ]{ Leibniz Institute for Crystal Growth, Max-Born-Stra\ss e 2, 12489 Berlin, Germany}

\begin{abstract}
Mixtures $(1-x)$ LiGdF$_4$ + $x$ LiLuF$_4$ ($0.50\leq x\leq0.75$) were melted and submitted to one zone melting cycle under a reactive HF atmosphere, aiming the study of the crystal growth of LiGd$_{1-x}$Lu$_x$F$_4$ solid solutions. Phase identification and compositional studies of the ingots were performed by scanning electron microscopy, energy dispersive X-ray spectrometry, and X-ray powder diffraction. Transparent regions of the bar related to the formation of LiGd$_{1-x}$Lu$_x$F$_4$ solid solutions were found to be enlarged proportionally to the initial amount of Lu in the mixtures. Gd segregates to the end of the ingot.
\end{abstract}


\end{frontmatter}


\section{Introduction}
\label{sec:intro}

Recently, some works reported the crystal growth and spectroscopic properties of LiLnF$_4$ (Ln = Lanthanides, or Y) mixed crystals \cite{Ranieri00c,Ranieri00,wetter02,Maldonado01}, aimed on the development of laser hosts with better optical and structural properties. The LiLnF$_4$ (LnLF) compounds are the intermediary phases formed in the binary systems LiF--LnF$_3$ described by Thoma \cite{Thoma70}, from Eu to Lu. These compounds crystallize in the scheelite structure, with $I\,4_1/a$ space symmetry group. The LiF--GdF$_3$ phase diagram was revised recently \cite{Ranieri04} and is characterized by the following data: a eutectic reaction at 25\,mol\% GdF$_3$ and $698^{\,\circ}$C and a peritectic reaction at 34\,mol\% GdF$_3$ and $755^{\,\circ}$C. For the LiF--LuF$_3$ system the phase diagram was revised by Harris \cite{Harris83} establishing: a eutectic reaction at 20\,mol\% LuF$_3$ and $704^{\,\circ}$C, a congruent melting point at 50\,mol\% LuF$_3$ and $850^{\,\circ}$C, and another eutectic reaction at 58\,mol\% LuF$_3$ and $832^{\,\circ}$C. The pseudo-binary section between the two scheelites GLF--LLF was investigated recently by thermal analysis \cite{Santos08}, and it was shown that scheelite mixed crystals under equilibrium conditions are crystallizing first from Lu-rich LiGd$_{1-x}$Lu$_x$F$_4$ melts $x\ge0.6$. In contrast, mixed rare earth trifluorides (Gd,Lu)F$_3$ are crystallizing first from Gd-rich melts, and will be converted upon cooling to scheelites Li(Gd,Lu)F$_4$ in a peritectic reaction with the LiF-rich melt. Under the crystallization conditions on the GLF--LLF section (Gd,Lu)F$_3$ crystallizes in the orthorhombic $\beta$-YF$_3$ structure with $P\,nma$ space symmetry group and shows complete miscibility for all compositions \cite{Ranieri08}.

The zone melting (ZM) technique is well known for its application in the purification of materials \cite{Roumie06} and in crystal growth as a relatively cheap and fast method \cite{Pfann52}. It is a useful tool to inspect the melting behavior of solid solutions, assuming complete miscibility in the liquid phase. From phases that melt incongruently a typical three region bar is obtained \cite{Abell76}, with the primary phase deposited at the beginning, the incongruent phase (here the scheelite solid solution) in the middle and a eutectic at the end.

This paper reports the growth of crystals from the LiGdF$_4$--LiLuF$_4$ (schee\-lite) section of the LiF--GdF$_3$--LuF$_3$ system by the ZM technique to inspect the melting behavior of the formed solid solutions. The different phases along the bars were investigated by scanning electron microscopy (SEM) including energy dispersive X-ray spectroscopy (EDS), and by X-ray powder diffraction (XRD) techniques.

\section{Experimental}
\label{sec:exp}

Samples were synthesized from commercial GdF$_3$ and LuF$_3$ (both 99.99\%, AC Materials) and LiF (99.9\%, Aldrich), previously purified by multiple zone melting in a flowing 1:1 mixture of hydrofluoric acid (HF) and argon (Ar) \cite{Guggenheim61} --- 4 cycles using a travel rate of 4\,cm/h and 2 subsequent cycles with 2\,cm/h. Syntheses  and zone melting processes were performed using  a platinum boat inserted in a sealed platinum reactor,  under this fluorinating atmosphere.  Four compositions of LiGd$_{1-x}$Lu$_x$F$_4$ ($x = 0.50; 0.60; 0.65$; and $0.75$) were melted in a furnace with an isothermal region of 220 mm lenght. For simplicity ingots are labeled by its nominal LiLuF$_4$ mole fraction as Lu50, Lu60, Lu65 and Lu75, respectively. Crystal growth was performed by single-pass zone melting these synthesized bars in a platinum boat (length 220\,mm, diameter 20\,mm), with a hot zone length of 2\,cm.
Initially, bars of Lu50 with masses of 150 and 85\,g were zone melted using travel rates of 2.5, 4.0 and 10\,mm/h. The best results were obtained for a travel rate of 4.0\,mm/h and a mass of 85\,g and were retained for all other compositions.

Cross sections of the bars were observed under a scanning electron microscope Philips XL30 with back-scattered electrons (BSE). Chemical compositions were estimated with an energy-dispersive spectrometer EDAX EDXAUTO. For the XRD analyses, the samples were pulverized to achieve a grain size around $38\,\mu$m and silicon was added as internal standard. The analysis was performed in the $2\,\theta$ range of 18$-70^\circ$. The lattice parameters and the molar fraction of the identified phases were calculated through the Rietveld method using the GSAS program \cite{Larson04}.

\section{Results and discussion}
\label{sec:results}

\begin{figure}[htb]
\centering
\includegraphics[width=0.46\textwidth]{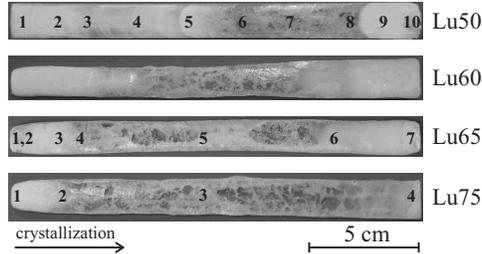}
\caption{LiGd$_{1-x}$Lu$_x$F$_4$ ingots obtained by zone melting technique. Lu50: zone rate of 2.5\,mm/h and mass of 150\,g; Lu60: zone rate of 4\,mm/h and mass of 85\,g; Lu65: zone rate of 4\,mm/h and mass of 85\,g; Lu75: zone rate of 4\,mm/h and mass of 85\,g. The number labels on the bars correspond to the numbers in the micrographs (Fig.~\ref{fig:SEM}).}
\label{fig:zm-bars}
\end{figure}

The four zone-melted bars are shown in Fig.~\ref{fig:zm-bars}. Three distinct regions can be identified in each of them: The first section to freeze is translucent or opaque due to its fine grained structure as it contains two phases. The following section is composed of large transparent grains of LiGd$_{1-x}$Lu$_x$F$_4$ solid solutions --- from here single crystalline scheelite grains can be harvested. The last section is opaque. It is obvious from Fig.~\ref{fig:zm-bars}, that increasing the Lu content in the matrix makes the transparent section wider. The share of this section in Lu50 is $\sim30$\% in all ingots processed, and is $\sim85$\% in Lu75.

\subsection{Microstructure characterization}

\begin{figure}[htb]
\centering
\includegraphics[width=0.47\textwidth]{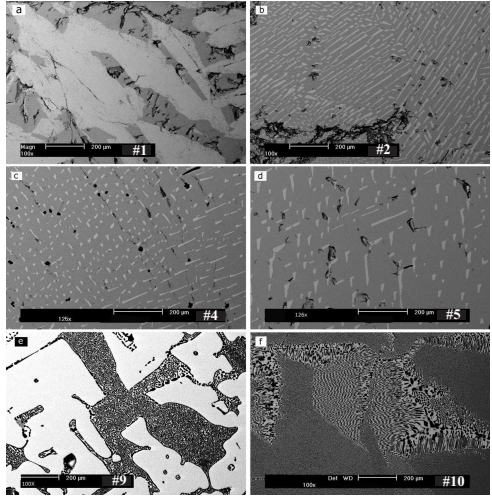}
\caption{SEM photographs (BSE contrast) in different regions of the ingot with Lu50 composition: a) first solidified part; b) to d) portion placed before the transparent region; e) just after the transparent region; f) last frozen portion.}
\label{fig:SEM}
\end{figure}

The typical microstructure for all ingots with Lu50 composition is shown in Fig.~\ref{fig:SEM}. These SEM images were taken from the ingot that is shown in Fig.~\ref{fig:zm-bars}. 

The typical microstructure for all ingots with Lu50 composition is shown in Fig.~\ref{fig:SEM}. These SEM images were taken from the ingot that is shown in Fig.~\ref{fig:zm-bars}. Two phases can be identified in the initial region of the ingots: a primary phase of random grains of (Gd,Lu)F$_3$, which is light gray in the micrographs a) to d); and darker regions composed by a Lu-rich solid solution of Li(Gd,Lu)F$_4$. The EDS data for each phase allowed to distinguish the (Gd,Lu)F$_3$ phase from Li(Gd,Lu)F$_4$ through the different fluorine concentration. The optically clear parts of the bar (sample points \#6, \#7, \#8) appear single phase also in the SEM micrographs (no contrast) and are therefore not shown in Fig.~\ref{fig:SEM}. 

In a sample selected just after the transparent portion, two phases are observed: a solid solution of Li(Gd,Lu)F$_4$ and a eutectic of LiF+Li(Gd,Lu)F$_4$ (darker region in the Fig.~\ref{fig:SEM}e). The microstructure of the last solidified part of the bar is mostly composed by eutectic colonies with disordered fibrous morphology (Fig.~\ref{fig:SEM}f). This disordered pattern is due to the occurrence of several simultaneous crystallization fronts. In general at the colony boundaries a coarse eutectic morphology is observed. The same microstructure was observed for the other ingots obtained with the Lu50 composition and it was independent of the parameters used in the ZM process. The Lu60 bar has a slightly higher lutetium concentration, but the observed microstructures are very similar to that of the Lu50 bars.

For Lu65, thermal instability in the beginning of the ZM process lead to a three phase microstructure, similar to an incomplete peritectic reaction: Li(Gd,Lu)F$_4$ phase is formed only on the (Gd,Lu)F$_3$ grain boundaries and there is a simultaneous crystallization of the eutectic. The last part of the Lu65 bar is characterized by the primary crystallization of Li(Gd,Lu)F$_4$ embedded in the eutectic. In the Lu75 composition bar, the instability at the beginning is smaller than for the Lu65 bar. In consequence, LiF is segregated forming the eutectic in a slight volume at the end of the bar. In the supplementary material one can find SEM images for the Lu65 and Lu75 bars. 

\subsection{Compositional characterization}

\begin{figure*}[htb]
\begin{center}
\includegraphics[width=0.95\textwidth]{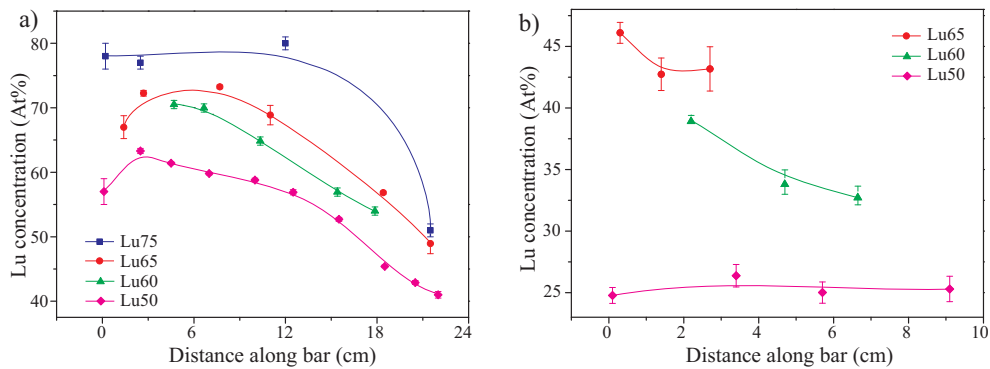}
\caption{Lu content along bars with 50, 60, 65 and 75\,mol\% Lu initial composition. a) for the Li(Gd,Lu)F$_4$ phase; b) for the (Gd,Lu)F$_3$ phase. The lines are only a guide for the eyes. Note different scaling of length axes.}
\label{fig:EDS}
\end{center}
\end{figure*}

The (Gd,Lu)F$_3$ trifluoride and the Li(Gd,Lu)F$_4$ scheelite are mixed crystals, and their chemical composition was estimated by EDS analysis without standard. Each phase was characterized with typically 6 independent measurements at different spots on the sample. The error bars in Figs.~\ref{fig:EDS} show the scatter of these measurements. Even though this technique is not appropriate to quantify the light element Li, it is quite adequate to measure the Ln elements.

Fig.~\ref{fig:EDS}b) shows that the (Gd,Lu)F$_3$ trifluoride phase crystallizes only in the first few centimeters of the bars, and this region becomes smaller for larger initial Lu concentration. For Lu75 (Gd,Lu)F$_3$ cannot be found in significant amounts that are suitable for analysis. Gd segregation in the (Gd,Lu)F$_3$ primary phases occurred in all cases, except for Lu50 where the trifluoride with constant composition of Gd$_{0.75}$Lu$_{0.25}$F$_3$ was crystallized along the first two-phase region.

Fig.~\ref{fig:EDS}a) shows that segregation in the Li(Gd,Lu)F$_4$ scheelite phase along the bars is strong for all initial concentrations. For the Lu75 sample, however, the initial composition LiGd$_{0.25}$Lu$_{0.75}$F$_4$ was nearly maintained in the first half of the bar. This is obviously due to the circumstance that the Gd-rich trifluoride phase (typically $>50$\% Gd, see Fig.~\ref{fig:EDS}b) does not crystallize first from such Lu-rich melts. Consequently, the only crystallizing phase scheelite has nearly the initial composition LiGd$_{0.25}$Lu$_{0.75}$F$_4$ --- this way allowing for a long stable crystallization process.

The EDS data permit to conclude that, if there are enough Gd ions available in the starting material, the trifluoride phase with an excess of Gd (close to Gd$_{0.75}$Lu$_{0.25}$F$_3$) crystallizes first. If the Gd concentration is too low, scheelite crystallizes first, and the initial Lu concentration in the scheelite is either equal to the initial concentration (for Lu75), or even higher then the initial concentration (Lu65, Lu60, Lu50), but then with strong segregation along the bar.

It is a general tendency for scheelite that Lu rich compositions tend to crystallize first, and subsequently the Lu concentration drops. This can be easily understood because both scheelite end members show unlimited miscibility and LiLuF$_4$ melts somewhat higher (congruently at $847^{\,\circ}$C) than LiGdF$_4$ (peritectically at $755^{\,\circ}$C) \cite{Santos08}. For the trifluoride the phase relations are more difficult: GdF$_3$ melts higher ($1252^{\,\circ}$C) compared to LuF$_3$ ($1182^{\,\circ}$C), but they show only limited miscibility \cite{Ranieri08}.

\subsection{Structural characterization}

\begin{figure*}[htb]
\begin{center}
\includegraphics[width=0.95\textwidth]{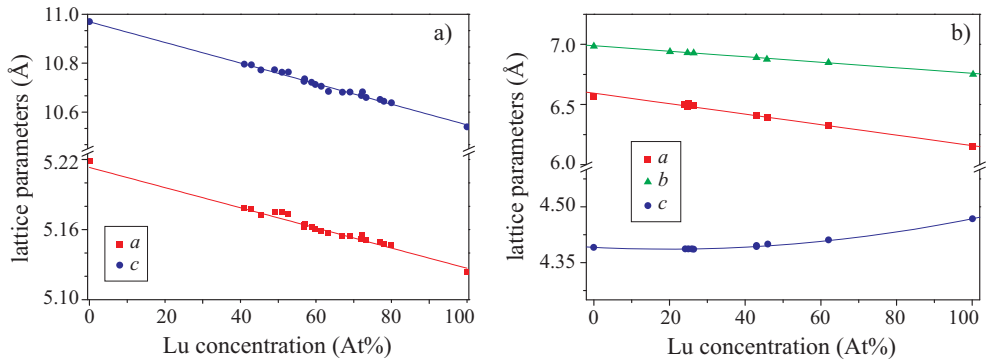}
\caption{a) Lattice parameters for the scheelite phase LiGd$_{1-x}$Lu$_x$F$_4$ vs. Lu concentration; b) Lattice parameters for the Gd$_{1-x}$Lu$_x$F$_3$ phase vs. Lu concentration. Additional data points for the end members $x=0$ and $x=1$ were obtained from separate measurements with the pure substances (see \cite{Ranieri08}).}
\label{fig:x-ray}
\end{center}
\end{figure*}

The identification of the phases present in each region of the bars was performed by X-ray powder diffraction and confirmed the EDS characterization. Lattice parameters and the share of the scheelite and trifluoride phases were calculated using the Rietveld method. (These results are presented in the supplementary material.) All the calculated lattice parameters of the Li(Gd,Lu)F$_4$ phase present in the ZM bars are plotted in Fig.~\ref{fig:x-ray}a), where the concentration values were obtained by EDS measurements. It can be seen that the incorporation of Lu in the mixed crystals decreases the lattice parameters for the scheelite phase proportionally.

Fig.~\ref{fig:x-ray}b) shows that the $a$ and $b$ lattice constants of the (Gd,Lu)F$_3$ trifluoride phase drop almost linearly if the Lu concentration becomes larger, whereas the $c$ parameter increases. The $c$ axis variation in this system is very similar to the $c$ axis variation of the pure rare earth fluorides in the whole rare earth group: in the orthorhombic rare-earth fluorides due to the lanthanides contraction, there is a linear drop of $c$ to 4.376\,\AA\ (for DyF$_3$) and then a strong rise up to 4.467\,\AA\ (for LuF$_3$) \cite{Shannon76}. Taking into account the mean ionic radius of the composition Gd$_{0.5}$Lu$_{0.5}$F$_3$, it is similar to the ionic radius of Ho$^{3+}$ that is 1.015\,\AA\ (with a coordination number of 8) \cite{Larson04,Shannon76}.

\section{Summary and Conclusions}

It could be demonstrated that the growth of scheelite type LiGd$_{1-x}$Lu$_x$F$_4$ solid solutions by zone melting is possible from mixtures of GLF and LLF, if the Lu concentration $x$ is sufficiently large, which means $x\gg0.5$. Figs.~\ref{fig:EDS}a) and \ref{fig:x-ray}a) show that then in the crystals the Lu concentration can range from $0.4\lesssim x\lesssim0.8$. However, segregation phenomena are typical during growth and result in concentration gradients, with lower Lu concentration in the later crystallized parts. For a starting concentration of 75\% LiLuF$_4$, however, the segregation was found to be small and LiGd$_{0.25}$Lu$_{0.75}$F$_4$ crystals were obtained for ca. 50\% of the floting zone bar length.

If the Lu concentration is small, orthorhombic Gd$_{1-x}$Lu$_x$F$_3$ crystallizes first, and the Lu concentration of this trifluoride phase also depends on the initial concentration and on the fraction of the material that is already crystallized. Gd$_{1-x}$Lu$_x$F$_3$ is not located on the section LiGdF$_4$--LiLuF$_4$ of the concentration triangle LiF--GdF$_3$--LuF$_3$. Consequently, free LiF remains in the molten zone. Besides, the Gd-concentration in the (Gd,Lu)F$_3$ phase is significantly higher compared to the initial concentration (Fig.~\ref{fig:EDS}b). This way the Lu-concentration of the molten zone grows until scheelite can crystallize. However, segregation proved to be strong for such Gd-rich initial compositions.

Depending on the Lu-concentration $x$ in LiGd$_{1-x}$Lu$_x$F$_4$ starting material, the following crystallization/melting behavior can be observed:

\begin{description}
\item[$x=0$:] GdF$_3$ crystallizes first at $857^{\,\circ}$C \cite{Klimm09b}. If the temperature goes below the peritectic line at $755^{\,\circ}$C, GLF is formed by the peritectic reaction with the LiF rich melt at the surface of the initial GdF$_3$ grains.
\item[$0<x\lesssim0.65$:] (Gd,Lu)F$_3$ solid solution crystallizes first, and the Gd concentration of this trifluoride is by ca. 30--50\% higher compared with the initial Gd concentration $1-x$. The bar length where (Gd,Lu)F$_3$ crystallizes first becomes smaller for larger $x$. Subsequently, scheelite Li(Gd,Lu)F$_4$ crystallizes. Upon cooling, the first crystallized (Gd,Lu)F$_3$ is partially converted by peritectic reaction with the LiF rich melt to scheelite: LiF + (Gd,Lu)F$_3\longrightarrow$ Li(Gd,Lu)F$_4$.
\item[$0.65\lesssim x<1$:] Already at the beginning mainly scheelite crystallizes, with minor contaminations by trifluoride only. The Gd segregation in the scheelite phase is considerably smaller, compared to the trifluoride phase. Nevertheless, there is segregation: the behavior remains incongruent (melt composition $\neq$ crystal composition), even if it is non-peritectic.
\item[$x=1$:] LLF crystallizes and melts congruently at $847^{\,\circ}$C \cite{Santos08}.
\end{description}

\section*{Acknowledgments}

The authors acknowledge financial support from FAPESP (05/57580-2), CNPq (477595/2008-1), and from CAPES - DAAD (368/11 and 50752632).

\section*{References}


\end{document}